\documentclass[useAMS,usenatbib]{mn2e}
\usepackage{graphicx}
\usepackage{aas_macros}

\title[Atmospheric monitoring with CASPER2]{Atmospheric monitoring in the mm and sub-mm bands for cosmological observations: CASPER2}

\author[M. De Petris et al.]
{M.~De Petris,$^1$\thanks{E-mail: marco.depetris@roma1.infn.it}
S. De Gregori,$^1$ B.~Decina,$^1$ L.~Lamagna,$^1$
J. R. ~Pardo,$^2$\\
$^{1}$Department of Physics, Sapienza University of Rome, Rome, I-00185, Italy\\
$^{2}$Centro de Astrobiolog$\acute{i}$a (CSIC/INTA), Instituto Nacional de T\'ecnica Aeroespacial, Madrid 28850, Spain\\}

\begin{document}

\date{Accepted 2012 November 7.  Received 2012 November 7; in original form 2012 October 1}

\pagerange{\pageref{firstpage}--\pageref{lastpage}} \pubyear{2012}

\maketitle

\label{firstpage}

\begin{abstract}
Cosmological observations from ground at millimetre and sub-millimetre wavelengths are affected by atmospheric absorption and consequent emission. The low and high frequency (sky noise) fluctuations of atmospheric performance imply careful observational strategies and/or instrument technical solutions.
Measurements of atmospheric emission spectra are necessary for accurate calibration procedures as well as for site testing statistics.
CASPER2, an instrument to explore the 90 $\div$ 450 GHz (3$\div$15 cm$^{-1}$) spectral region, was developed and verified its operation in the Alps. A Martin-Puplett Interferometer (MPI) operates comparing sky radiation, coming from a field of view ($fov$) of 28 arcminutes (FWHM) collected by a 62-cm in diameter Pressman-Camichel telescope, with a reference source. The two output ports of the interferometer are detected by two bolometers cooled down to 300 mK inside a wet cryostat. Three different and complementary interferometric techniques can be performed with CASPER2: Amplitude Modulation ($AM$), Fast-Scan ($FS$) and Phase Modulation ($PM$). An altazimuthal mount allows the sky pointing, possibly co-alligned to the optical axis of the 		2.6-m in diameter telescope of MITO (Millimetre and Infrared Testagrigia Observatory, Italy). Optimal timescale to average acquired spectra is inferred by Allan variance analysis at 5 fiducial frequencies.
We present the motivation for and design of the atmospheric spectrometer CASPER2. The adopted procedure to calibrate the instrument and preliminary performance of the instrument are described. Instrument capabilities were checked during the summer observational campaign at MITO in July 2010 by measuring atmospheric emission spectra with the three different procedures.
\end{abstract}


Ground-based cosmological observations can be carried out in the millimetre and sub-millimetre bands (hereafter mm and sub-mm) but need a continuous monitoring of the atmospheric contribution. Accuracy in the calibration procedure towards known photometric sources results from a good knowledge of transmission and emission along this optical path.
Sky brightness can be suitably monitored in this frequency range by spectrometers designed only with this goal (see as example the FTS interferometers employed in \cite{Matsuo98}, \cite{Matsushita99}, \cite{Paine00}) or, more generally, to spectral observations of sky objects when optically matched to existing telescopes (see the description of the instrument installed at the CSO focal plane in \cite{Serabyn96}).
We have developed an instrument, CASPER2, to record atmospheric emission spectra in the millimetre band for assisting cosmological observations with the 2.6-m in diameter Cassegrain telescope at MITO (Testa Grigia, Italy, 3480 m a.s.l.) (\cite{DePetris07}). It is conceptually similar to CASPER, the experiment proposed for the italian-french Antarctic base, Dome C (\cite{DePetris05}), but with a more limited spectral band (\cite{Decina10}).
Such an instrument makes it possible to avoid specific telescope procedures, with consequent loss of observational time like skydips, and to infer atmosphere opacity in a wide spectral range.\\
After a brief description of the importance of an instrument like CASPER2 in Sect. 2, the instrument concept and its major characteristics are described in Sect. 3. In Sect. 4 a discussion about the Martin-Puplett interferometer (MPI) and the options adopted for signal sampling are presented.
Sect. 5 reports preliminary atmospheric spectra as recorded at MITO in July 2010. Final remarks are discussed in the Conclusions.
\section{Cosmological observations requirements}
Ground based observations, focused on cosmological targets in the mm-submm bands (sub-THz frequencies), are greatly affected by atmospheric presence. The performance in this spectral region is mainly dominated by water vapour content modeled by a continuum-like term and by a series of absorption lines peaked at 183, 325, 380, 448, 557 GHz. Oxygen is also present as absorber at 118, 368 and 425 GHz but, differently from H$_2$O fluctuations, it is well-mixed in the atmosphere and so it mainly contributes only as a constant radiation background, still dependent on the elevation, with consequent photon noise on the detectors. Atmospheric synthetic spectra, derived from models, are a valuable support for predicting atmosphere brightness starting from thermodynamic parameter values, see for example the ATM code (\cite{Pardo01}). However a validation of the model for a specific site is mandatory and this is possible only when on-site recorded spectra are available. This is what is going on at MITO with CASPER2.\\
Short term variation of the atmospheric emission on different spatial scales contributes as sky noise while slow fluctuations of the attenuation can affect the quality of sky calibrations. Regarding only observations in the millimetre/sub-millimetre bands, two kinds of approaches for atmosphere monitoring are widely applied. The atmosphere is usually dealt with by an isothermal slab model where opacity follows a zenithal secant-law. Under this assumption large telescopes are periodically employed to perform altitude scans in the sky, ${\it i.e.}$ skydips, to derive the zenith optical depth, integrated in the operating photometric band. Due to the fact that this procedure subtracts time from the astronomical observations, the opacity is alternatively derived by employing ancillary instruments such as tippers, GPS or water vapour radiometers (see for instance \cite{Radford98}, \cite{Coster96} and \cite{Wiedner01}). All these usually feature wide fields of view and limited spectral capabilities, unmatched to those of scientific instruments. This makes the assessment of the effective site quality (i.e. in-band transparency and fluctuations at the angular scales of interest) a non-trivial task.
A spectrometer is the natural solution for a continuous sky monitoring over a wide spectral band and an MPI is an efficient solution for reaching this goal. Furthermore if the spectrometer is optically matched to an its own medium size telescope it is possible to observe the sky towards a specific direction and at an angular scale closer to that of the main telescope.
In any case a model for extrapolating the atmosphere opacity at different frequencies is mandatory.\\
A semi-empirical procedure was proposed by \cite{Degregori12} to perform an analysis of atmospheric transmission employing radiosoundings data to feed the ATM code for generating synthetic spectra in the wide spectral range from 100 GHz to 2 THz.\\

\section{Instrument}
CASPER2 is composed of a two-mirror telescope, an interferometer, a wet cryostat and an altazimuthal mount.
The atmospheric emission spectra are acquired using an MPI that can be easily converted into a spectropolarimeter with many consequent scientific applications. Here the MPI is fully described, while in Sect. 3 the basics of the interferometer and the calibration procedures are detailed. The detectors, two Ge-bolometers with a NEP $\sim 10^{-15}$ W Hz$^{1/2}$, are cooled down to 290 mK by a wet cryostat, using liquid nitrogen and helium combined with a He$^{3}$ fridge. The cryostat (Infrared Labs, HDL-8) is identical to the one described in \cite{Catalano04}. In Figure \ref{figure1} a schematic CAD drawing of CASPER2 is shown while Figure \ref{figure2} illustrates the instrument in operation at the Testa Grigia station (Alps, 3480 m a.s.l.).\\
The optical design satisfies the expected measurement requirements: a low resolution spectrometer in the range of 90 - 450 GHz
(R $\sim$ 50) and a small field of view (less than 1 degree). To achieve these goals we matched a two-mirror telescope with an MPI. The optical layout with the main components, as labelled in the text, is shown in Figure \ref{figure3}.\\
A f/3.5 Pressman-Camichel telescope collects sky radiation and feeds one of the two input ports of the MPI. With this optical solution the wavefront aberrations introduced by the 62-cm concave spherical primary mirror ($M1$) are compensated by a 12-cm convex ellipsoidal subreflector ($M2$). The primary mirror is underilluminated by the secondary mirror, which operates as an aperture stop of the telescope alone resulting in an entrance pupil 46-cm in diameter. Both the mirrors were manufactured in an aluminum alloy, ensuring a fast and homogeneous thermalization of the surfaces. The mirrors were carefully polished to reflect visible light: the final $rms$ surface roughness is lower than 0.1 $\mu m$. The primary mirror has been manufactured by Officine Ottico Meccaniche Marcon di San Don\'a di Piave (Italy) while the ellipsoidal subreflector by the INFN machine workshop at the Department of Physics in Rome.
The secondary mirror is maintained in the right position along the telescope axis by a 2 cm thick polystyrene foam plate ($ST$) (BASF, Styrodur 3035N), allowing a null obscuration due to the subreflector support also avoiding the possible consequent diffracted radiation. Laboratory measurements show
high transmission value ($>$ 97 per cent) and low polarization ($<$ 1 per cent)  of this material in the whole band.
The telescope is shielded from off-axis unwanted radiation ($i.e.$ Sun during daytime measurements) by 8 panels with inner reflective surfaces, 0.5 mm thick Peraluman sheets, shaped as vanes to operate like roof mirrors (\cite{Gervasi98}).

\setcounter{figure}{0}
\begin{figure}
\centering
\includegraphics[width=8.0cm]{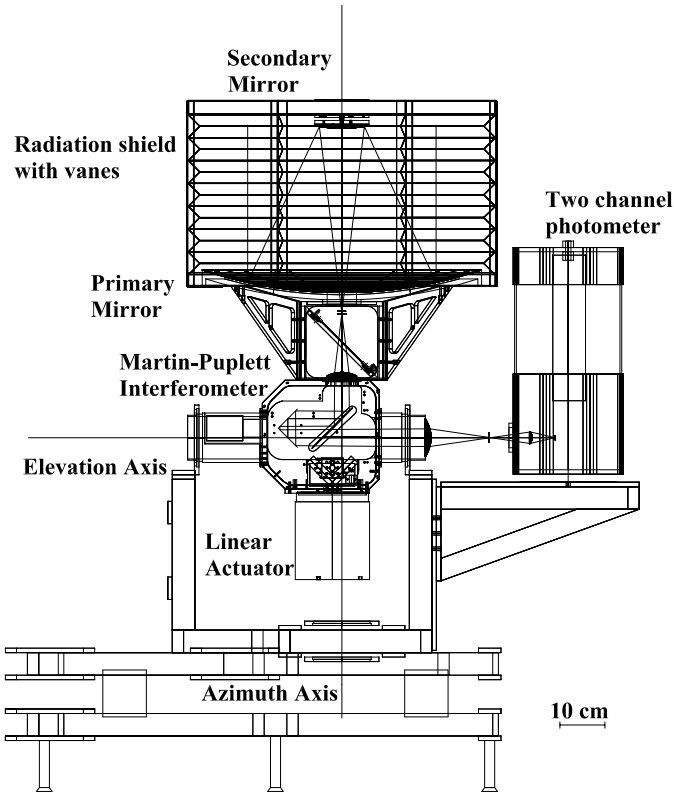}
\caption{CAD drawing of CASPER2.}
\label{figure1}
\end{figure}

\setcounter{table}{0}
\begin{table}
\caption{CASPER2 instrumental main features.}
\label{table0}
\centering
\begin{tabular}{l c}
\hline
Telescope                								  &  f/3.5 Pressman-Camichel  					\\

Telescope effective focal lenght          &  1621 mm                            \\

Primary mirror diameter   								&  620 mm            									\\

Primary mirror conic constant							&  0                                  \\

Primary mirror curvature radius					  &  978.4 mm														 \\

Secondary mirror diameter   						  &  120 m             									\\

Secondary mirror conic constant					  &  8.86                               \\

Secondary mirror curvature radius         &  354.9 mm                           \\

Entrance pupil diameter                   &  460 mm															\\

Field of view (FWHM)          					  &  26 arcmin  												\\

$A\Omega$                 								&  0.05  $cm^2$ sr   									\\

Interferometer          								  &  Martin-Puplett 										 \\

Spectral range           									&  Channel 1: 90 $\div$ 360 GHz       \\
                          								&  Channel 2: 90 $\div$ 450 GHz       \\
                          								
Mechanical path difference								&  30 mm															 \\

Spectral resolution       								&  5 GHz															 \\

Detectors                                 &  2 composite NTD                    \\
																					&  bolometers @ 0.3 K                 \\

Calibrator                                &  Eccosorb AN72                      \\

Mount                                     &  altazimuthal	                      \\

Star tracker field of view                &  1 arcmin                           \\

CCD field of view                         &  (14.4 x 13.6) arcmin               \\
\hline
\end{tabular}
\end{table}

\subsection{Optics}\label{optics}

A 45 degrees tilted wire-grid ($WG1$) linearly polarizes the transmitted sky radiation focused by the telescope ($sky$) (see Fig. \ref{figure3}). The radiation emitted by a reference source at ambient temperature ($ref$), realized by a disc of Eccosorb AN72, enters, as the second input port of the MPI, reflected by $WG1$ with a perpendicular polarization. An alternative colder reference source, closer to atmospheric emission, is under consideration. The two inputs are combined in planar waves, at the entrance of the MPI, by an HDPE (High Density PolyEthylene) plano-convex lens ($L1$). A second wire-grid ($WG2$), still tilted at 45 degrees inside the MPI, is rotated around the optical axis of 35.26 degrees to correctly split the two polarization axes. It acts as beam splitter separating the linearly polarized incoming radiations into two orthogonal components. Two 90 degree roof-mirrors ($RM1$ and $RM2$), diamond-turned stainless steel, can move along the two split optical beams changing the Optical Path Difference (OPD). $RM1$ is mounted on a linear stage (AICOM S.P.A., Model SMP-123) to perform a $\pm$15 mm mechanical path difference, corresponding to a 5 GHz spectral resolution, while $RM2$ is sinusoidally wobbled by a shaker (Lynge Dynamic System, Model 409) on a linear stage with an amplitude of approximately 1 mm (see Sect.\ref{PM}).

The radiation undergoes a polarization rotation of 90 degrees when reflected back onto the roof mirrors. A second HDPE plano-convex lens ($L2$) focuses on the beam, exiting from the MPI, in front of the cryostat window. The last HDPE plano-convex lens ($L3$) is mounted inside the cryostat, on the radiation shield of the helium liquid tank, cooled down to 1.6 K. This lens refocuses the radiation towards two detectors, i.e. the two output ports, after the splitting of the two polarization states by the third wire-grid ($WG3$), cooled down to 300 mK. We name Channel 1 the port corresponding to the transmitted radiation, and Channel 2 the other. All the wire-grid polarizers have 10-$\mu m$ in diameter tungsten wires spaced by 25-$\mu m$ and show an efficiency in the reflected and transmitted polarizations better than 10$^{-3}$ in our spectral range.

The first two lenses, L1 and L2, were carefully shaped to image the subreflector on $L3$, at least for the Zero Path Difference (ZPD) $RM1$ position, defining it as the cold aperture stop of the full optical system, $i.e.$ the Lyot stop.  We have the possibility of selecting the last optical element as aperture stop only because a single pixel is present at the focal plane. Incidentally $L3$ is also the exit pupil. The optical design was developed with ZEMAX\footnote{ZEMAX Development Corporation; www.zemax.com} ensuring diffraction limited performance in the whole 90 $\div$ 450 GHz spectral range.
The two bolometers are illuminated by f/3.5 Winston cones with a 10.5 mm in diameter aperture entailing a throughput of 0.05 cm$^2$ sr.

The optical axis, from the MPI exit to the cryostat entrance window, overlaps the elevation axis, ensuring a vertical position for the cryostat for all the telescope pointing positions. The optical consequences of this choice, as derived from the variable orientation between the wire grids with the elevation angle, are discussed in Sect. 3.

After fixing the telescope configuration and assuming a $fov$ equal to 28 arcminutes (FWHM) we limited the lower value for the focal length of the lens $L1$, $f_{L1}$, to satisfy the Jacquinot condition. The minimum spectral resolution can be related to telescope-MPI optical matching in the following way:

\begin{equation}
\delta \nu  \geq \frac{\nu_{max} }{8}  \left( \frac{d_{tfs} }{f_{L1}} \right)^2=
\frac{\nu_{max} }{8}  \left( \frac{fov f_{tele} }{f_{L1}} \right)^2
\end{equation}

where $\nu_{max}$ is the maximum frequency (450 GHz), $d_{tfs}$ the telescope field stop diameter (13.2 mm) and $ f_{tele}$ the telescope effective focal length (1621 mm).
In our case $f_{L1}$ was chosen equal to 145 mm satisfying the Jacquinot condition up to R$\sim$1000 at $\nu_{max}$.

\setcounter{figure}{1}
\begin{figure}
\centering
\resizebox{\hsize}{!}{\includegraphics{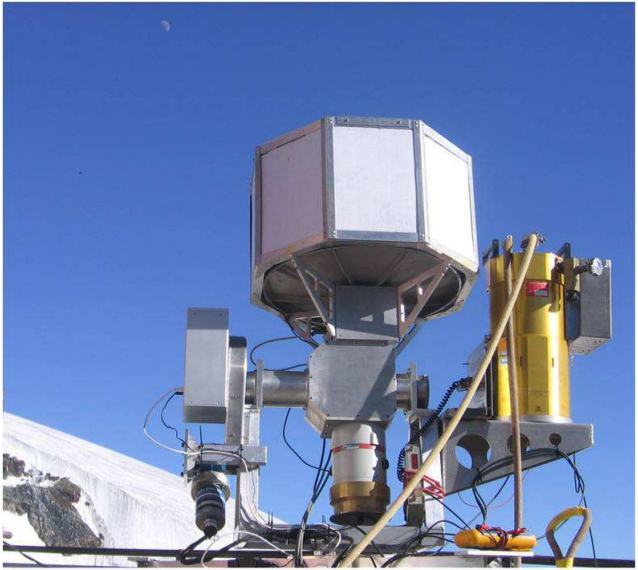}}
\caption{CASPER2 in operation at Testa Grigia station (3480 m a.s.l.) close to MITO telescope during the July 2010 observational campaign.}
\label{figure2}
\end{figure}

\setcounter{figure}{2}
\begin{figure}
\includegraphics[width=8.2cm]{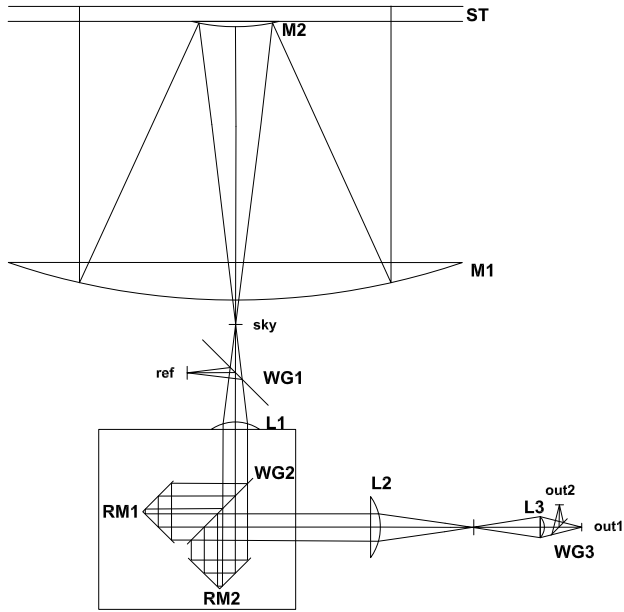}
\caption{CASPER2 optical layout: all the labeled components are described in the text.}
\label{figure3}
\end{figure}

\subsection{Filters chain}
Blocking filters are employed in our system in order to reduce
radiative input during the different cryogenic stages in the cryostat
(and consequently on the bolometers) while mesh filters to select the
frequency bandwidth of interest. In Table \ref{table1} all the components in the filters chain are listed. We chose to perform measurements inside two similar bandwidths: 90$\div$360 GHz
for Channel 1 and 90$\div$450 GHz for Channel 2.
Channel 2 was spectrally enlarged to explore high frequency atmosphere emission more prone, and so more sensitive, to $pwv$ fluctuations at the expense of a larger background emission with consequent photon noise: 0.5 nW and 1 nW  respectively, under the assumption of a $pwv$ = 1 mm.

\setcounter{table}{1}
\begin{table}
\caption{ Filters chain characteristics}
\label{table1}
\centering
\begin{tabular}{c c c}
\hline
type                           & cut\ (GHz) & Temperature (K)\\
\hline
Quartz window                  & off ($>$ 3000)    &300\\
\hline
ARC quartz +                   & off ($>$ 1200)   &77\\
black polyethylene                                    \\
\hline
Yoshinaga                      & off ($>$ 1650)     &1.6\\
Mesh                           &off ($>$ 450)      &1.6\\
\hline
Yoshinaga                      &off ($>$ 1500)      &0.3\\
Mesh channel 1                 &off ($>$ 360)      &0.3\\
Mesh channel 2                 &off ($>$ 450)      &0.3\\
Winston cones                  &on ($<$ 90)        &0.3\\
\hline
\end{tabular}
\end{table}

The vacuum window of the cryostat is a 4 mm thick quartz window.
The 77 K quartz filter is 3.3 mm thick with a diamond powder ARC.
A black polyethylene filter, 0.1 mm thick, is employed to reduce visible and NIR background. At 1.6 K a Yoshinaga filter blocks IR radiation while an interference mesh filter limits the high frequencies at 450 GHz.
On the entrance apertures of the two Winston cones at 300 mK, two
Yoshinaga filters work as last IR blockers. Two final mesh filters perform the effective band selection. All the filters were supplied by QMC Instruments, Ltd. The lower limit band in frequency is dictated by the output aperture of the Winston cones operating as high-pass filters.

\subsection{Electronics and data acquisition}
The first stage for the read out of the bolometer signals are JFETs (model IFN146) in a common-drain configuration. Those unitary gain amplifiers are mounted on the He$^{4}$ tank inside an aluminum shielded box, close to the detectors, but at a temperature of 120 K by self-heating. The measured noise is $\sim$ 3 nV/$\sqrt{Hz}$. The expected total incident power on the detectors is 0.5 nW and 1 nW  respectively, implying a sensor thermal conductivity of the order of $10^{9}$ W/K (Haller-Beeman).
The low impedance signals from the JFETs feed an ambient temperature differential preamplifier with a gain equal to 250 for both the channels and an output noise $\sim$ 10 nV/$\sqrt{Hz}$ for frequency higher than 20 Hz.
Rechargeable batteries inside the preamplifier box supply the bias voltage, as well as all the read out electronics boards and cryogenics maintenance. All the electrical connections are ensured by twisted pairs of NbTi wires (0.1 mm diameter) shielded within CuNi (0.03 mm thick).
Depending on the signal modulation (see Sect. \ref{modulation}) the data acquisition is carried out by a low frequency ($<$1 Hz) data sampling of two synchronous demodulation amplifiers (Stanford Research Instruments, SR850) or a high frequency (5 kHz) data sampling by an ADC (National Instruments PCI-6031).

\subsection{Pointing System}

The pointing system was developed with a twofold aim. The first is to track in the sky the field of view of the MITO
telescope during its observations, exploring in this way an identical optical path through the atmosphere, while the second goal is to be able to point specific directions to perform skydips for every azimuth angle. The telescope control is realized  with Magellano ST7, supplied by ATEC Robotics (Advanced Technologies for Research and Industry). The movement of the two axes is performed by hybrid stepper motors (MAE, HS200) while the position is transduced by two incremental encoders (Baumer Electric, BHK 16OSA400). Both encoders ensure an accuracy of
400 steps/turn and they have a zero position to record and to reset the sky coordinates every observational run. This reference position also corresponds to the telescope rest position.

The gear ratio for the altitude (and azimuth) axis is equal to 5.1 employing two gear wheels with 112 and 22 teeth respectively. In order to have stable and accurate movements, this low value was increased
by a right angle gear (TLS, Sf40/PB3) with a 100:1 gear ratio.
Finally the total gear ratio is 510. The transmission is ensured
by two gear belts (Trasmecam, HTD 1040-8M-20 for altitude axis and
1200-8M-20 for azimuth axis).

A CCD camera (SBIG, ST-402ME) is used to check the pointing system, as star tracker. It is provided with a 135 mm focal length lens to have a total field of view of 14.4' x 13.6', with a field of view per pixel of about 14". The limit magnitude is 15.

The co-alignment between visible and millimetre axes was checked in the laboratory with a fixed artificial source, considering its finite distance, and than verified \emph{in situ} by pointing planets. The final error in pointing is lower than 1 arcminute,
more than enough for atmospheric emission measurements.

The whole instrument is accommodated inside a 1.6x2.9x1.4 m (WxDxH) deployable dome in PVC Pre\'contraint \textregistered \  502 (AMA, series 8000).

\section{The Martin-Puplett interferometer}

The MPI employs, as polarizers, three wire-grids distributed inside the instrument as described in Sect. \ref{optics}.
First of all we recall the basics of an MPI.

\subsection{MPI basics}
\label{mpi}
Stokes and Mueller matrix formalism can be used to describe any state of polarization and superposition of beams when no phase relation has to be taken into account. Alternatively, as for example in the analysis of an interferometer, one needs to perform a Jones matrix calculation, reverting to the Mueller matrix formalism afterwards (\cite{Catalano04}).

We assume unpolarized radiation at the entrance of the two input ports and we describe it with the Stokes formalism as

\begin{equation}\label{S_I}
S_{in}= B_{in}\left(
\begin{array}{c}
1  \\ 0 \\ 0  \\ 0
\end{array}\right)
\end{equation}

where $in$ stands for $in_1$ and $in_2$, the two input sources.

The two signals pass through $WG1$ and are linearly polarized in the following way:

\begin{equation}\label{S'}
S'_{in}= M_{WG1} \cdot S_{in}=B_{in}\left(
\begin{array}{c}
1  \\ \cos2\vartheta_{in} \\
\sin2\vartheta_{in}
\\ 0
\end{array}\right)
\end{equation}

where $\vartheta_{in}$ is the projected angle of the $WG1$ principal axis on the plane orthogonal to the optical axis.
The angles for the two inputs are linked as: $\vartheta_{in_2} = \vartheta_{in_1} + \pi /2$.
$\vartheta_{in_1}=0$ degrees corresponds to the vertical position.
The Muller matrix used for a wire grid is

\begin{equation}\label{MWG}
M_{WG}= \left(
\begin{array}{cccc}
1 & \cos2\vartheta_{in} & \sin2\vartheta_{in} & 0 \\ \cos2\vartheta_{in} & \cos^{2}2\vartheta_{in} & \cos2\vartheta_{in}\sin2\vartheta_{in} & 0 \\ \sin2\vartheta_{in} & \cos2\vartheta_{in}\sin2\vartheta_{in} & \sin^{2}2\vartheta_{in} & 0
\\ 0 & 0 & 0 & 0
\end{array}\right)
\end{equation}

After the MPI, each exiting beam can be described as:

\begin{equation}\label{S"}
S^{''}_{in}=M_{MPI} \cdot S'_{in}=B_{in}\left(
\begin{array}{c}
1  \\ \cos2\vartheta_{in}\cos\delta \\
-\sin2\vartheta_{in}
\\ \cos2\vartheta_{in}\sin\delta
\end{array}\right)
\end{equation}

where we used the Muller matrix for an ideal MPI

\begin{equation}\label{MPI}
M_{MPI}= \left(
\begin{array}{cccc}
1 & 0 & 0 & 0 \\ 0 & \cos\delta & 0 & \sin\delta \\ 0 & 0 & -1 & 0
\\ 0 & \sin\delta & 0 & -\cos\delta
\end{array}\right)
\end{equation}

The mechanical path difference between the two split beams  is $\Delta x_{mec}$.
The OPD, $\Delta x_{opt}$, equal to $2 \Delta x_{mec}$, is related to the phase shift, $ \delta$, for each wavelength, $\lambda$, as well known: $ \delta = 2 \pi \Delta x_{opt}/\lambda $.

$WG3$ splits each beam in the output ports in the following way:

\begin{equation}
\label{S'''}
\begin{array}{c}
S^{'''}_{o}=  \\ M_{WG3} \cdot S^{''}_{in}= \\
 B_{in}\left(
\begin{array}{c}
1 + \cos2\vartheta_{in}\cos2\varphi_{o}\cos\delta -
\sin2\vartheta_{in}\sin2\varphi_{o} \\ \cos2\varphi_{o} +
\cos2\vartheta_{in}\cos^{2}2\varphi_{o}\cos\delta -
\sin2\vartheta_{in}\sin2\varphi_{o}\cos2\varphi_{o} \\
\sin2\varphi_{o} + \cos2\vartheta_{in}\sin2\varphi_{o}\cos2\varphi_{o}\cos\delta -
\sin2\vartheta_{in}\sin^{2}2\varphi_{o}
\\ 0
\end{array}\right)
\end{array}
\end{equation}

where $o$ stands for $out_{1}$ or $out_{2}$, corresponding to the two output ports optically matched to Channel 1 and Channel 2, respectively.

Due to the fact that our detectors are only sensitive to signal intensity, $i.e.$ the
first element of Stokes vector, we can express the signals of the two output ports as:

\begin{equation}\label{ch1}
I_{out_{1}}= \Delta^{+} + \Delta^{-}\cos2\vartheta_{in_1}\cos2\varphi_{out_{1}}\cos\delta -
\Delta^{-}\sin2\vartheta_{in_1}\sin2\varphi_{out_{1}}
\end{equation}

\begin{equation}\label{ch2}
I_{out_{2}}= \Delta^{+} - \Delta^{-}\cos2\vartheta_{in_1}\cos2\varphi_{out_{1}}\cos\delta +
\Delta^{-}\sin2\vartheta_{in_1}\sin2\varphi_{out_{1}}
\end{equation}

where $\Delta^{+}=B_{in_1}+B_{in_2}$ and $\Delta^{-}=B_{in_1}-B_{in_2}$ and rewriting $\varphi_{out_{2}} = \varphi_{out_{1}} + \pi/2$.

Under the assumption that $\vartheta_{in_1} = \varphi_{out_{1}}$, $i.e.$ the polarization axis of $WG1$
is aligned with the $WG3$ one, we can rewrite Eqs.
\ref{ch1} and \ref{ch2} in the following way:

\begin{equation}\label{Ch1final}
I_{out_{1}}=\Delta^{+} + \Delta^{-}\cos\delta
\end{equation}

\begin{equation}\label{Ch2final}
I_{out_{2}}=\Delta^{+} - \Delta^{-}\cos\delta
\end{equation}

The spectra of the two input sources are linked to Eqs. \ref{ch1} and \ref{ch2} in Sect. \ref{calib}.

\subsection{MPI efficiency \emph{versus} telescope altitude}

The cold $WG3$, inside the cryostat, is at rest during telescope altitude movements. Its wires change orientation by an angle $\varphi_{out_{1}}$ with respect to the input polarizer $WG1$, linked to the altitude angle $\alpha$ as $\phi_{out} = \theta_{in_1} + \alpha -\frac{\pi}{2}$.
For this reason we refer to the function $f(\alpha)$, as an instrumental efficiency, the signal dependence with the altitude. This has
to be carefully taken into account to characterize the pointing performance.
The two outputs at ZPD, see Eqs. \ref{ch1} and \ref{ch2},  can be expressed as:

\begin{equation}\label{out12}
I_{out_{1,2}}=\Delta^{+} \pm \Delta^{-} f(\alpha)
\end{equation}

where $f(\alpha) = cos 2 \alpha $.

When $\alpha=45$ degrees the signals, even modulated by $\cos\delta$ along the
interferogram, are null.
To check this expected instrument efficiency we filled the sky input with a blackbody source at ambient temperature (an Eccosorb AN72 sheet in front of the telescope) and, in order to get a high signal-to-noise ratio, we employed a Hg-lamp as reference source instead of the disc of Eccosorb AN72. Spectra were acquired at altitude angles ranging from -10 to 100 degrees. We also explored altitude angles lower than 0 degrees and higher than 90 degrees to check the consistency and the repeatability of the performance.
Interferograms were recorded in fast scan mode (see Sect. \ref{FS}) with a 5 kHz scan rate, performing the time average over an acquisition time of about 7 minutes  for altitude values ranging from 30 degrees to 60 degrees (lower instrumental efficiency), and of 5 minutes for the others (higher instrumental efficiency).
In Figure \ref{figure4} the two  ZPD output signals are plotted versus the altitude angle, with error bars given by the standard deviation.
At zenith, Channel 1 is proportional to $\Delta^{+} + \Delta^{-}$, after an intensity decreases down to a null signal ($\Delta^{-} = 0$), the rotation of polarization axis inverts the signal at horizontal position. The red points represent the polarized signal detected by Channel 2, as expected the trend is symmetrical to the other. Data points are well-fitted by the  $\pm cos 2 \alpha$ function predicted by Eq. \ref{out12} (see the residuals in the bottom panel of Fig.\ref{figure4} )

\setcounter{figure}{3}
\begin{figure}
\centering
\resizebox{\hsize}{!}{\includegraphics{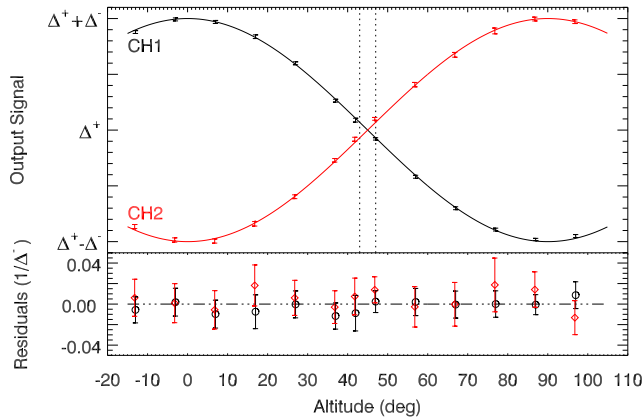}}
\caption{Dependence between ZPD output signals and altitude angle.
The two vertical dashed lines limit the "$blind$" observational range where the signal to noise ratio is lower than 3.
In the bottom panel the normalized signals with the instrument function removed are shown (black circles for Channel 1 and red diamonds for Channel 2).  }
\label{figure4}
\end{figure}

The loss of efficiency, due to the change of $WG$s orientations along different zenithal angles, entails an altitude range where the signals are negligible.

In order to determine the width of this altitude range where the instrument has low efficiency, what we call the "$blind$" observational range, we estimated $\alpha_{min}$ and
$\alpha_{max}$ corresponding to a signal-to-noise ratio $\leq 3$. An emission atmospheric spectra, as derived by ATM with a $pwv$ = 1 mm, was assumed as source. The vertical dot-dashed lines in Figure \ref{figure4} limit this "$blind$" orientation directions: 43 degrees $<\alpha<$ 47 degrees.

This is a restriction on the CASPER2 performance but it has few consequences for our purposes because the minimum altitude explored by the MITO telescope is 42 degrees. This operational limit for MITO telescope is dictated by low atmospheric contamination requirements and by dome constraints. However CASPER2 can point lower altitudes, where the signal-to-noise ratio increases, allowing a complete angular range skydips.

The expected variation in the atmospheric emission during skydips, due to the decreasing opacity with the altitude, has to be added to the instrumental efficiency.
Each detector records a signal that can be expressed as:

\begin{equation}\label{signal}
I_{out_{1,2}}=\Delta^{+}(\alpha) \pm \Delta^{-} a(\alpha) f(\alpha)
\end{equation}

where $a(\alpha)$ is the atmospheric dependence and $\Delta^{-}$ is the input signals difference at the zenith position. We can notice that, due to the different spectral bands, $\Delta^{+}$ and $\Delta^{-}$ have to be distinguished between the two channels. In a simple optically-thin atmospheric layer model we can assume

\begin{equation}\label{alpha}
a(\alpha)=exp(-\csc(\alpha))
\end{equation}

normalized at the zenith position.

\subsection{The sub-interferometer: Mickey}\label{Mickey}

A Michelson sub-interferometer, named Mickey, ensures the monitoring of the movements (position and velocity) of $RM1$ during fast scan measurements. The movable mirror is mounted on the back side of $RM1$. The source is a laser (Imatronic, Sigma 650/3) centered at $\lambda$=650 nm.
Interference maxima are detected by a photodiode (Osram, SFH203) and then processed by a peak-counter circuit. Accuracy on the mirror position depends on the distance between the peaks generated by Mickey, equal to $\lambda$=325 nm.

The Fourier Transform of the signal detected by Mickey along a return double-sided interferogram shows two maxima, corresponding to a backward and forward scan velocity difference of about 70 $\mu $m/s.
The peak-counter circuit is employed to produce a trigger signal for data collection. In this way all fast scan interferogram points are acquired at the same distance between each other, independently of velocity changes due to the step motor.

\subsection{Thermal monitoring}
An AD590 temperature sensor is used to monitor the absolute temperature of the reference load. The reference source passively follows the ambient temperature as well as the whole instrument.
The MPI box and the primary mirror are also monitored to estimate possible differential emission in the instrument due to temperature gradients. The AD590 is a temperature transducer
producing an output current proportional to absolute temperature, suitable for our purposes:
wide temperature range (-55$^\circ$C $\div$ 150$^\circ$ C), high calibration accuracy ($\pm$
0.5$^\circ $C) and excellent linearity ($\pm$ 0.3$^\circ$C over full range).

\subsection{Signal modulation techniques} \label{modulation}
The peculiarity of CASPER2 is the ability to perform, with the same interferometer, 3 different signal modulations: Amplitude Modulation ($AM$), Fast-Scan ($FS$) and Phase Modulation ($PM$).
The OPD variation along the time for the three options is shown in Fig \ref{fig7}.
Many fast-scanning FTSs are built so that the ZPD position lies close to one edge of the moving stage path, in order to maximize the dynamic range
of OPD and consequently the spectral resolution. Anyway, since one-sided interferograms transform into real spectra, no information on the
phase is available, although phase problems do show up as baseline anomalies. Two-sided interferograms, on the contrary, transform
into complex spectra (they have two pieces of information per frequency), allowing phase errors to be directly measured as a function of frequency.
This permits checking for optical mis-alignments and other potential instrumental problems through the level of asymmetry in the two sides of
the interferograms. Since a high spectral resolution is not necessary when measuring the continuum level of the atmospheric emission, CASPER2
adopts a two-sided interferogram sampling, with the ZPD located half-way along the moving mirror path.
To reduce the effect of measuring the interferograms with a limited mechanical path difference, the interferogram lobes are weighted with a triangular apodization function, at the expense of decreasing the spectral resolution to about 8.6 GHz.

\subsubsection{Amplitude Modulation}\label{AM}

The amplitude modulation ($AM$) technique is fulfilled by a chopper wheel modulator placed in front of the cryostat, performing a synchronous demodulation of the signal, and a step-by-step movement of $RM1$.
The step length is equal to 100$\mu m$ ensuring a Nyquist frequency, $\nu_{N}$ = 750 GHz, higher than 450 GHz, see Table \ref{table1}.
The chopper blades, coated with Eccosorb AN72, act as a reference source at ambient temperature. The two modulated signals feed the two lock-in amplifiers.
Since the chopper is kept at room temperature, the large brightness temperature gradient between the 2 sources implies that the faint signal fluctuations are difficult to detect. In addition the $AM$ procedure has the disadvantage that half of the observation time is spent on the reference source.
The $AM$ procedure represents a reliable tool useful for characterizing the performance of the whole instrument, working both with stable sources in laboratory tests and instrument calibration over bright sky signals.

\subsubsection{Fast-Scan}\label{FS}

The fast-scan technique consists in sweeping the range of available  OPDs through a rapid movement of the translating stage at constant velocity.
The recorded time-domain signal is therefore trivially related to the interference pattern thanks to a simple time/position conversion through the stage velocity, and the Fourier analysis yields directly the necessary information about the spectrum  of the incoming radiation beam.
A carefully selected velocity may shift the electrical frequencies of interest away from potentially troublesome low-frequency components or line features in system noise, allowing for a cleaner reconstruction of the optical power in the passband.
Moreover, since the scan can be repeated an arbitrary number of times, the short integration time per unit OPD can be increased to hit the photon noise limit with almost no additional effort in the instrument setup.
On the other hand, apart from mechanical limitations, an intrinsic upper limit to the value of the translation velocity is determined by the time constant of the detectors and by the highest frequency in the instrument bandwidth: in order to be able to discriminate two consecutive fringes in
the interference pattern generated by radiation at frequency $\nu_{max}$, they must be scanned in a time interval longer than the
detector time constant $\tau_d$, thus determining the limit velocity

\begin{equation}
v_{lim} \leq \frac{1}{\tau_d \nu_{max}}
\end{equation}

Fast scanning interferometers are usually operated well below this limit, and signals are de-convolved from the detector time-response before
further data-processing to avoid residual artifacts. This is also the case for CASPER2, where the time-constant of the detectors is $\tau_d$=10 ms and
the translation velocity of the moving roof mirror is set to 0.86 cm s$^{-1}$ ($i.e.$ 1.72 cm$^{-1}$ lag velocity).
Under these conditions, a time of 3.8 s is needed to perform one scan. An average of several scans on the same source is needed to improve the signal to noise ratio of the observation (see Sect. \ref{Allan}).

A critical source of systematics in fast-scanning FTSs is the non-uniformity of the scan velocity: since interferograms need to be
uniformly sampled in space-domain, rather than in time-domain, any fluctuation in velocity at a fixed scan rate converts into a local "stretch" of the OPD scale, resulting in artifacts in the frequency domain, both in line and continuum interferometry.
In CASPER2, this issue is solved by monitoring the stage position through the optical sub-interferometer (see Sect. \ref{Mickey}), whose fringes provide a position reference for each data-point.
The interferograms are highly oversampled (5 kSamples s$^{-1}$), with an electrical Nyquist frequency of 2.5 kHz (or an equivalent optical bandwidth of 43.6 THz). After Fourier transformation, the high frequency components of instrument noise are discarded and the signal in the optically meaningful band is processed to extract the
sky brightness and $pwv$ information.

One of the advantages of fast scan interferometry over step by step interferometry (or slow scan) is that no chopper is used to modulate the radiation (see Sect. \ref{AM}).
A disadvantage in $FS$ interferograms is that slow drifts in the intensity of the source can result in variations of the baseline of the interferogram which can be of the same frequency as modulations from the longest wavelength being measured. This affects the performance at low optical frequencies.

\subsubsection{Phase Modulation}\label{PM}

A third signal modulation technique, available with CASPER2, is the Phase Modulation ($PM$).
This technique, proposed by \cite{Chamberlain71a}, allows the replacement of $AM$ and $FS$ implying a
modulation of the OPD when recording the interferogram.
The insensitivity to slow fluctuations in the intensity of the source plus the full time on source observations, make $PM$ very attractive to atmospheric measurements.

Operatively modulation is performed by periodically wobbling $RM2$ while the detector signals are lock-in demodulated.
We can rewrite the Muller matrix for the MPI, as in Eq.\ref{MPI} :

\begin{equation}\label{newmp}
M_{MPI}= \left(
\begin{array}{cccc}
1 & 0 & 0 & 0 \\ 0 & \cos\delta^{'} & 0 & \sin\delta^{'} \\ 0 & 0
& -1 & 0
\\ 0 & \sin\delta^{'} & 0 & -\cos\delta^{'}
\end{array}\right)
\end{equation}

where $\delta^{'}=\delta+\delta_{M}$ with $\delta_{M}=2\pi
M(t)/ \lambda$, the modulation of the OPD obtained by periodically oscillating $RM2$
with an amplitude $A$ and a frequency $\nu_{0}$. In Figure
\ref{fig7} the OPD along the time is plotted for all the three observation modes.

\setcounter{figure}{4}
\begin{figure}
\centering
\resizebox{\hsize}{!}{\includegraphics{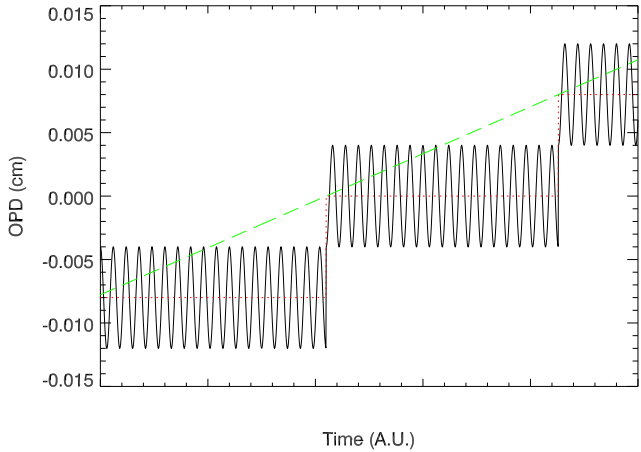}}
\caption{Comparison of variation of OPD along the time
for $FS$ (green dashed line), $AM$ (red dotted line) and $PM$ (black solid line) modulation techniques.}
\label{fig7}
\end{figure}

The amplitude of the phase modulation has to be carefully chosen to fit the spectral band of the instrument.
In our case, with a sinusoidal modulation function, $A = 0.58 c/ \nu_{max}$ that for $\nu_{max}$ = 450 GHz, the highest frequency in the two bands, corresponds to 840 $\mu m$.

The wobbling frequency is chosen equal to 12 Hz so that the effect of low
frequency source variations is essentially eliminated using $PM$
techniques. At the same time, by avoiding a $dc$ component the acquisition dynamic range is well fitted to the interferogram range.

$RM2$ is wobbled by a linear actuator driven by a waveform generator using a feedback loop based on a position transducer and a Proportional-Integral-Derivative circuit. A linear variable differential transformer {\it (LVDT)} (Solartron Metrology, model SM/1) is used as position transducer for the mirror.
The accuracy in oscillation amplitude and frequency was checked and found to be 2 per cent in amplitude and 1 per cent in frequency.
The requirements on the stability of the oscillation amplitude were estimated by simulated observations. Changes in the amplitude result in a different weighting along the spectrum affecting the inferred $pwv$, as derived by fit of ATM synthetic spectra. The constraint on $pwv$  respecting the uncertainty on the $RM2$ oscillation amplitude turns out to be 2 per cent, estimated assuming several $pwv$ average values in the range 0.1 mm to 6 mm.

\subsection{Calibration procedures}\label{calib}
Calibrated spectra are derived by employing several considerations.
The two inputs of CASPER2 differ as follows:

\begin{equation}
B_{in_1}(\nu) = \epsilon_{atm}(\nu)BB(T_{atm},\nu) + \epsilon_{tele}(\nu)BB(T_{tele},\nu)
\end{equation}

\begin{equation}
B_{in_2}(\nu) = BB(T_{ref},\nu)
\end{equation}

where $\epsilon_{atm}$ and $\epsilon_{tele}$ are the atmospheric and telescope emissivities respectively while $BB$ stands for the specific brightness of the atmosphere, the telescope and the reference load (the only one having an emissivity equal to 1), each of them at the equivalent temperatures $T_{atm}, T_{tele}$ and $T_{ref}$.
While $\epsilon_{atm}$ is related to the atmospheric opacity, we assume $\epsilon_{tele}$ equal to $3 \cdot 10^{-3}$ at the frequency of 150 GHz, for an aluminum mirror, and changing with the frequency as $\sqrt{\nu}$ (\cite{Bock95}).

The spectra derived from Eqs. \ref{ch1} and \ref{ch2}, after baselines removal, are related to the spectra of the incoming sources as:

\begin{eqnarray}
\tilde{I}_{out_1}(\nu) & = & R_{1}(\nu) \varepsilon_{1} (\nu) A\Omega (\nu)  \left[ B_{in_1}(\nu) - B_{in_2}(\nu) \right] = \nonumber \\
               & = & F_{1}(\nu) \left[ B_{in_1}(\nu) - B_{in_2}(\nu) \right]
\end{eqnarray}

\begin{eqnarray}
\tilde{I}_{out_2}(\nu) & = & R_{2}(\nu) \varepsilon_{2} (\nu) A\Omega (\nu) \left[ B_{in_2}(\nu) - B_{in_1}(\nu)\right] = \nonumber \\
               & = & F_{2}(\nu) \left[B_{in_2}(\nu) - B_{in_1}(\nu)\right]
\end{eqnarray}

The calibration function, $F_{i}(\nu)$, for the $i$-channel includes the responsivity $R_{i}(\nu)$, the spectral efficiency $\varepsilon_{i} (\nu)$ and the throughput $A\Omega (\nu)$. In Fig. \ref{figure5bis} the normalized calibration functions are shown for both channels.

\setcounter{figure}{5}
\begin{figure}
\centering
\resizebox{\hsize}{!}{\includegraphics{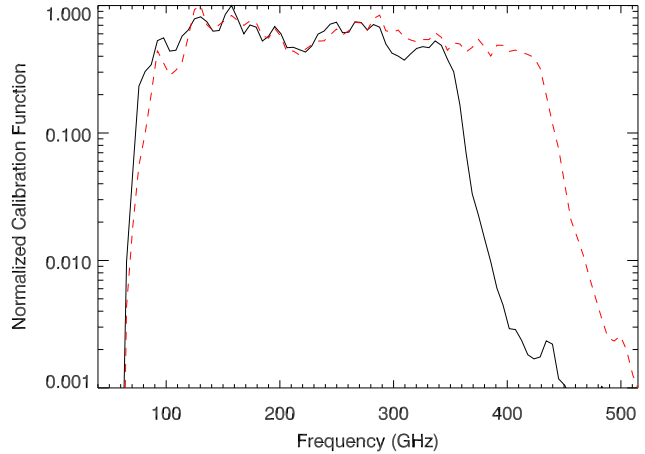}}
\caption{The normalized calibration function for CH1 (black solid line) and for CH2 (red dashed  line).}
\label{figure5bis}
\end{figure}

The throughput $A\Omega (\nu)$ is assumed equal for the two inputs, being only dictated by the optical matching between the cones and the last cold lens operating as aperture stop. A frequency dependence of it can be considered for both patterns: starting from a single mode propagation, at the cone exit apertures at the lowest frequency, and moving to a multi mode approach at 450 GHz.
For CASPER2's application this anisotropic response of the two ports is not a technical hitch due to the presence of diffuse sources, the atmosphere and the sheet of Eccosorb, totally filling both the inputs.
We can easily assume that the telescope efficiency (\cite{Ruze66}) is unitary along the whole spectral range counting on an $r.m.s.$ surface error of $\sim 0.1 \mu m$ and that the wire grids behave like ideal polarizers in our bands.

The optical matching between the MPI and the two sources shows differences due to the distinct optical paths. Specifically in the case of $in_1$, the input port is further transformed by the primary beam telescope.

The calibration functions are estimated by filling the $in_1$ port with a second well modeled source, a cold black body, realized with an Eccosorb AN72 sheet thermalized inside a liquid nitrogen bath (77 K). Hence the atmospheric spectra is deduced by:

\begin{equation}
I_{atm_i}(\nu) = \frac{I_{out_i}(\nu)}{F_{i}(\nu)} + BB(T_{ref},\nu) - \epsilon_{tele}(\nu)BB(T_{tele},\nu)
\end{equation}

Possible optical mismatch between the two input ports has to be known and taken into account to avoid
the consequent contamination on inferring the atmospheric spectra.
The correct balance between the two inputs is
checked by filling even the sky port with a room temperature blackbody inserting an
Eccosorb AN72 sheet in front of the telescope to record a {\sl null} interferogram.
Only in this case a residual signal could arise in the acquired spectrum when a far from ideal source coupling of
the two input ports is present. A potential temperature gap between the two blackbodies could also produce a non-{\sl null} interferogram but this however has been monitored.

The knowledge of the {\sl null}  interferogram, or at least the upper limit of the ZPD value when the signal-to-noise ratio is less then one, allows us to put a constraint on the minimum detectable contribution on the $pwv$ content.
Long acquisition of {\sl null} interferograms enable us to discriminate spectra with a difference of only 0.01 mm of $pwv$ irrespective of the $pwv$ content, at least for $pwv < $ 1 mm.

\subsection{The Allan Variance}\label{Allan}
The instability of the atmospheric emission in the mm/sub-mm spectral region has to be carefully taken into account when the observational goal is to achieve frequent and independent high signal-to-noise ratio spectra. The time dedicated to performing a single interferogram is mainly dictated by the detector time response while the timescale to average several interferograms is affected by the atmospheric drifts.
In the specific case of the $FS$ technique, to avoid loss of observational time, the minimum number of spectra that can be averaged has to be constrained achieving a high signal-to-noise ratio while  sampling a continuous changing atmosphere. It is fundamental to determine a characteristic time to indicate when an instrument is dominated only by thermal noise instead of an atmosphere fluctuations regime. An appropriate approach to infer this timescale is to estimate the Allan variance (\cite{Allan66}). In a wide band spectrometer, like CASPER2, it is  important to also check  the timescale similarity for all the frequencies by investigating the noise performance in the measured spectra (\cite{Schieder01}).

Large fluctuations of atmosphere emission are expected in correspondence with the 3  "windows" centered at the frequencies of 150, 270 and 350 GHz; hereafter we refer to these lines with ${\bf 2}$, ${\bf 4}$ and ${\bf 5}$, respectively. On the contrary, the oxygen band at 118 GHz and the high absorption $H_2O$ band at 183 GHz, quoted as ${\bf 1}$ and ${\bf 3}$, should appear more stable with time. In Fig. \ref{figure7}, for example, 87 spectra of the zenithal atmospheric brightness recorded by the $FS$ technique are shown for the 2 bands of CASPER2. Each spectra is the average of a couple of back-and-forth spectra acquired in 6 seconds. The vertical dotted lines refer to the examined frequencies.

The Allan variance is calculated, for the previous 5 reference lines, in the following way. The signal, $s_f(t_i)$, related to the atmospheric emission at the frequency $f$ extracted from the $FS$ spectra at time $t_i$, is averaged over variable timescales, $T$, generating the new dataset:

\begin{equation}
S_f(T,t_j) = \frac{1}{T} \sum^{j+T}_{i=j+1} s_f(t_i)
\end{equation}

The Allan variance, or the two-sample variance, is estimated for the frequency $f$ as in \cite{Wiedner02}:

\begin{equation}
{\sigma^{2}_A}(T) = \frac{1}{N-2} \sum^{N-1}_{j=2} \left(\frac{S_f(T,t_{j-1}) + S_f(T,t_{j+1})}{2} - S_f(T,t_j) \right)^2
\end{equation}

In Fig.\ref{figure8} the Allan variance corresponding to the 5 frequencies is plotted in the cases of the two CASPER2 bands. It is worth noting the expected $1/T$ dependence for short average times, corresponding to dominant thermal noise, and the slope change when atmospheric drifts overcome. This behaviour is not satisfied in the case of ${\bf 1}$ and ${\bf 3}$, where a strong and stable emission is present.

We can estimate, at least for this specific dataset of spectra, $T \simeq 100$ s as the more suitable average time. Even if the values of the Allan variance increase for all the frequencies in the high-background Channel 2, affected by a larger instrumental noise, the best average time is almost the same. We employ this timescale to generate the spectra reported in the next Section.

\setcounter{figure}{6}
\begin{figure}
\centering
\resizebox{\hsize}{!}{\includegraphics{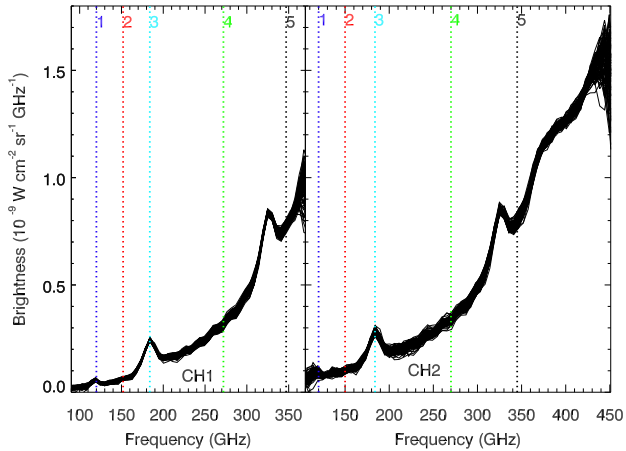}}
\caption{Overplotted atmospheric zenithal spectra recorded at MITO on July 16th 2010, starting at 04:03 AM and finishing at 04:12 AM. The vertical dotted lines denote the fiducial frequencies, defined in the text, where the Allan variance has been estimated. }
\label{figure7}
\end{figure}

\setcounter{figure}{7}
\begin{figure}
\centering
\resizebox{\hsize}{!}{\includegraphics{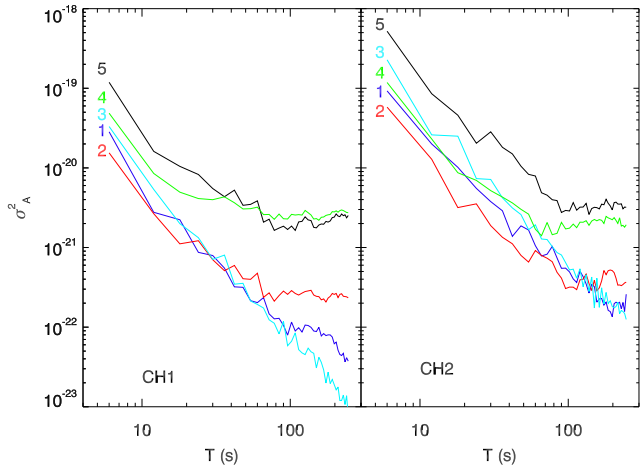}}
\caption{Allan variance estimated at the 5 fiducial frequencies labeled and colored as in Fig. 7.}
\label{figure8}
\end{figure}

\setcounter{figure}{8}
\begin{figure}
\includegraphics[width=6.2cm,angle=90]{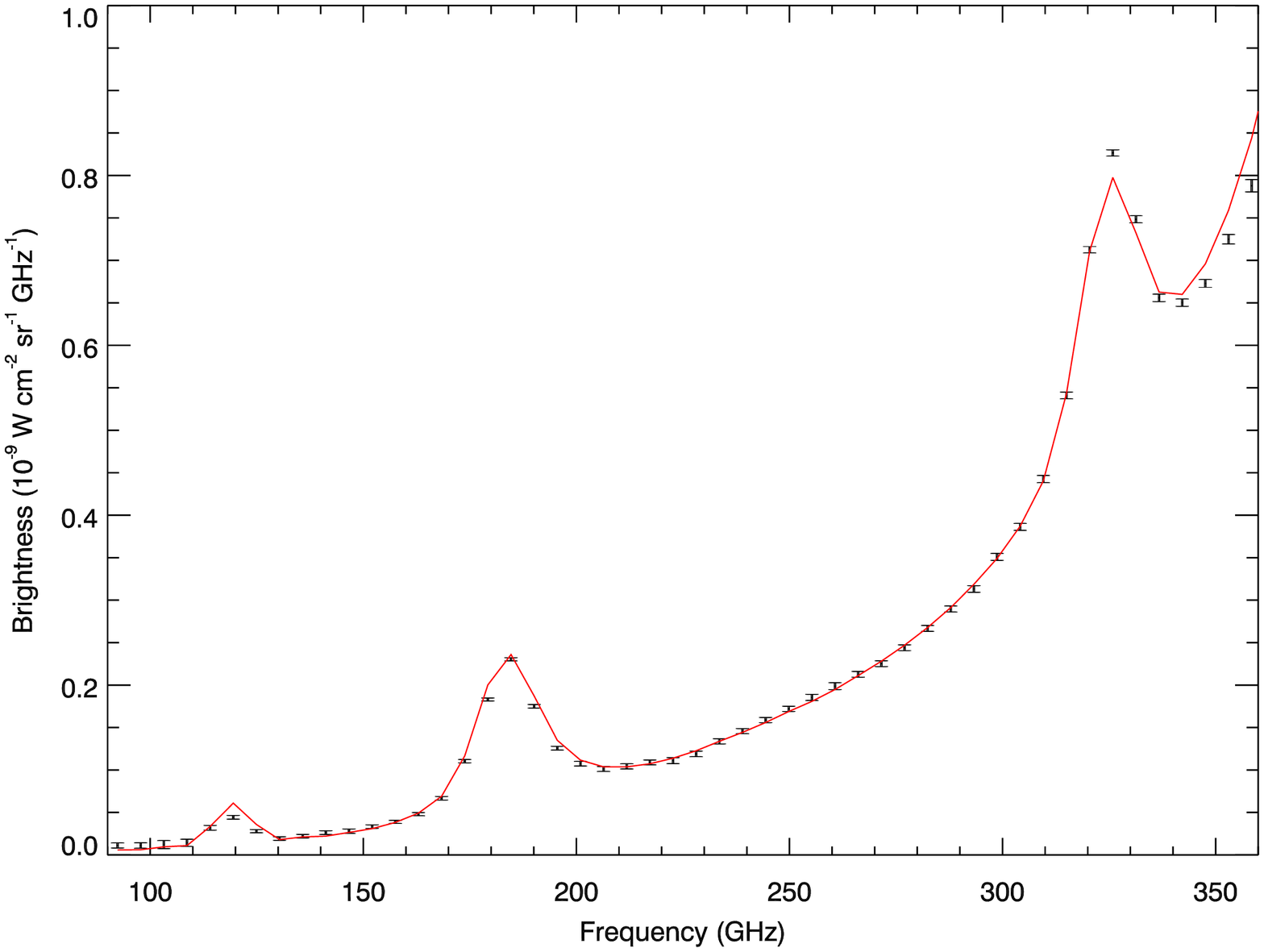}
\includegraphics[width=6.2cm,angle=90]{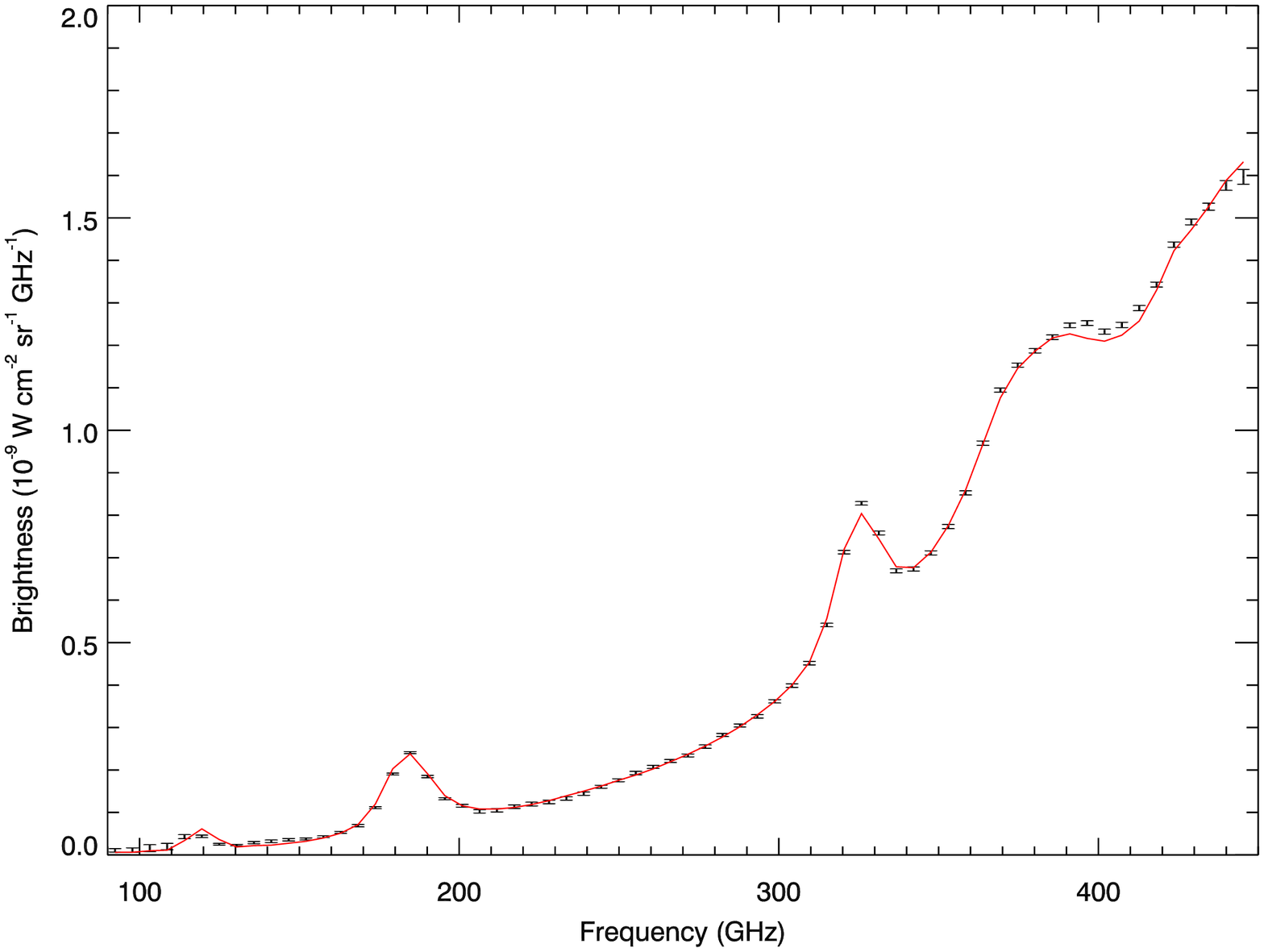}
\caption{Averaged atmospheric zenithal spectra recorded with fast scan procedure at MITO on July 16th 2010, 04:03 AM. The red line is the best fit obtained with ATM model corresponding to $pwv = 6.53 \pm 0.16 $ mm for CH1 (top panel) and  $pwv = 6.84 \pm 0.16$ mm for CH2 (bottom panel).}
\label{fig9}
\end{figure}

\section{Preliminary atmospheric spectra measurements at MITO}\label{Spectra}
In this section we present preliminary measurements of atmospheric spectra performed during the summer campaign at MITO from 11th to 19th July 2010.
In order to show the instrument's capabilities, atmospheric spectra measured by the two channels of CASPER2 with fast scan procedure on July 16th 2010 are shown in Fig. \ref{fig9}.
The uncertainty associated with the data points is given by the standard deviation of the dataset and it is closely connected to the time on which the interferograms average is performed (see Fig. \ref{figure8}).
Best fit ATM model corresponds to $pwv$= 6.53 $\pm$ 0.16 mm for CH1 and $pwv$= 6.84 $\pm$ 0.16 mm for CH2 (red lines in Fig. \ref{fig9}).
The uncertainty on the $pwv$ value was evaluated by a random generation of the ATM synthetic spectra within the brightness uncertainty range and it corresponds to less then 3 per cent.
Missing of consistency between data and ATM spectra with increasing the frequency, supports the necessity to accurately calibrate the dry continuum and the $H_2O$ pseudocontinuum terms in the simulated atmosphere (\cite{Pardo01}). The ATM model is currently based on T/P typical profiles and the validation of the code is the subject of a subsequent paper.

\section{Conclusions}

A double beam FTS interferometer, installed at the focal plane of a 62-cm in diameter telescope, is devoted to monitoring atmospheric emission spectra in the mm/sub-mm band compared to an ambient temperature reference source.
The MPI is designed to perform three independent signal acquisition procedures: fast scan, phase modulation and step and integrate.
Being an ancillary instrument, it is provided by an independent altazimuthal mount allowing to point the same direction of a telescope dedicated to cosmological millimetre observations from the ground; in the current version the 2.6-m in diameter MITO telescope in the Alps. The recorded spectra, in the 90$\div$450 GHz spectral region, permit to validate the results of the transfer radiative code, ATM (\cite{Pardo01}), for this site and consequently to
infer the $pwv$ value as derived by fit with synthetic spectra or by skydips.
The instrument has been characterized in laboratory and employed during the observational campaign at MITO in July 2010.
The choice of the best spectra integration time is derived by the Allan variance estimate.
Preliminary spectra recorded by fast scan modulation are presented to show the instrument's capabilities.

\section*{Acknowledgments}
Part of this work was supported by Sapienza University of Rome with Ateneo 2009 (C26A09FTJ7) and 2011 (C26A11BYBF). We acknowledge with gratitude F. Melchiorri's suggestions which were a valid contribution during the early phases of this project. Useful proposals were received by G. Pisano, G. Siringo, G. Savini, E.S. Battistelli, P. de Bernardis and E. Caca. We also thank the referee for the constructive comments which have improved the presentation of our results and S. Capaldi for the English revision of the manuscript. Mechanical and a few optical components of the instrument have been realized at the machine-shop of the Department of Physics and at Officina Meccanica INFN - Sezione di Roma. Preliminary observations were performed at MITO with the logistic support provided by Sezione IFSI/INAF di Torino, now Osservatorio Astrofisico di Torino.

\bibliographystyle{mn2e}
\bibliography{mybib}

\label{lastpage}
\end{document}